\documentclass[pre,aps,twocolumn,superscriptadress,floatfix, 10pt]{revtex4-1}

\usepackage[utf8]{inputenc} 
\usepackage[T1]{fontenc}   
\usepackage[english]{babel} 

\usepackage{placeins}
\usepackage{hyperref}       
\hypersetup{colorlinks=true, urlcolor=cyan}

\usepackage{latexsym}
\usepackage{silence}
\usepackage{soul}
\usepackage{dsfont}
\usepackage{lipsum}

\usepackage{setspace}    
\usepackage{indentfirst} 
\usepackage{titlesec}    

\usepackage[margin=2cm]{geometry} 

\usepackage{array}
\usepackage{color}

\usepackage{multirow} 

\usepackage{amsmath}
\usepackage{amssymb}
\usepackage{amsfonts}
\usepackage{amsthm}
\usepackage{cancel}
\usepackage{bbold}

\usepackage{empheq}

\usepackage{bbm}
\usepackage{bm}

\usepackage{braket}
\usepackage{slashed}

\usepackage{algorithm}
\usepackage{algpseudocode}
\usepackage{cleveref}

\usepackage{graphicx}
\graphicspath{{./Assets/}}

\usepackage{tikz}
\usetikzlibrary{calc,shapes.geometric,arrows}

\usepackage{wrapfig}
\usepackage{ragged2e}
\makeatletter
\long\def\@makecaption#1#2{
  \par\footnotesize
  \begingroup
  \@parboxrestore
  \@setminipage
  \justifying
  #1.~#2
  \par
  \endgroup
}
\AddToHook{cmd/appendix/before}{\def\cref@section@alias{appendix}}
\makeatother

\definecolor{darkgreen}{HTML}{008000}

\begin{document}

\title{Identical Particle Systems : Hierarchical Spectral Reconstruction}

\author{Hovan Lee}
\email{hovan_l@protonmail.com}
\affiliation{Department of Physics, Royal Holloway, University of London, TW20 0EX, Egham, United Kingdom}
\author{R\'{e}mi Lef\`{e}vre}
\affiliation{Department of Physics, Royal Holloway, University of London, TW20 0EX, Egham, United Kingdom}
\author{Gr\'{e}goire Ithier}
\affiliation{Department of Physics, Royal Holloway, University of London, TW20 0EX, Egham, United Kingdom}

\date{\today}

\begin{abstract}
We identify a hierarchical symmetry structure underlying the Hilbert space of quantum systems made of identical particles.
By studying the linear map between single-body and many-body spectra, we show that the resulting spectral organization admits
a natural partition into sectors induced by a symmetry governed by cyclotomic fields and their Galois groups.
This approach offers a new perspective on the structure of quantum many-body spectra, defining a controlled coarse-graining
of the Hilbert space in which degeneracies and spectral features can be systematically organized through a set of new quantum
numbers. The resulting structure induces a hierarchy of spectral resolutions following a renormalization flow, which enables
information loss control.
We demonstrate that this hierarchical decomposition provides an efficient route to approximate MBDoS calculations for any
identical particle system, while preserving physically relevant spectral properties. This approach provides a general
symmetry-based framework for organizing and approximating many-body spectra, with potential applications to quantum
thermalization, spectral statistics, and large-scale quantum simulations.
\end{abstract}

\maketitle

\section{Introduction}

\label{sec:introduction}

The exponential growth of Hilbert space dimension in quantum many-body systems presents a fundamental barrier to the
direct computation and interpretation of spectral properties. 
Even in non-interacting systems, the many-body density of states (MBDoS) inherits a combinatorial complexity arising from the rapid
growth of the occupation-number configuration space, leading to an intractable enumeration of many-body configurations for all but
the smallest system sizes.
Understanding how physically relevant spectral information is organized within this exponentially large space remains an open
challenge in quantum statistical mechanics and quantum information theory.

Standard approaches to this problem include exact diagonalization \cite{EDweisse2008exact}, stochastic sampling techniques \cite{bethe1936attempt},
recursive methods \cite{jacquemin1986exact}, kernel polynomial expansions \cite{silver1994densities,silver1996kernel,KPM-RevModPhys.78.275,sobczyk2022spectral},
and tensor network methods \cite{PhysRevB.96.094303}.
While powerful, these methods primarily address the computational problem of spectral estimation, rather than revealing an
underlying structural organization of the many-body spectrum itself. As a result, degeneracies and combinatorial redundancies
are typically treated as computational obstacles rather than manifestations of deeper symmetry principles.

In this work, we propose a different viewpoint : the many-body spectrum of identical particles possesses an intrinsic hierarchical
structure governed by symmetry-induced sectors of the underlying occupation space.
By analyzing the singular value decomposition structure of the matrix of occupation numbers we allow system-specific quantities to
be isolated away from intrinsic combinatorial properties \cite{lefevre2023many}, then identify a decomposition of many-body
configurations into a family of symmetry sectors characterized by cyclotomic fields and their Galois groups.
These symmetry sectors can be labeled by new quantum numbers that are compatible with a hierarchic
organization of many-body configurations : they define a natural notion of coarse-graining in which we identify a renormalization
flow \cite{lefevre2026symmetries}.

The renormalization flow of sectors induces a hierarchic representation of the MBDoS, ranging from exact resolution to progressively
compressed approximations with decreasing computational costs : this can be used to selectively truncate high resolution sectors
to meet accuracy and performance requirements adaptively.
Importantly, this truncation is not ad hoc but arises from an intrinsic ordering of symmetry information in the many-body configuration
space. Within this framework, computational efficiency emerges as a consequence of a deeper organization of spectral degeneracies.

We demonstrate how this symmetry-induced representation enables controlled approximations of the MBDoS for any identical particle
systems, and show that it captures the emergence of thermal-like distributions from purely combinatorial structure.
Beyond its computational implications, the framework suggests a broader interpretation of many-body spectral statistics in terms of
number-theoretic symmetry structures, providing a new lens for studying quantum many-body systems.


\section{Framework Foundations}

\subsection{The Hilbert-Space Lifting Map}



We begin by introducing the central object of our framework and define the Hilbert-space lifting map as :
\begin{equation*}
    F_{\hat{O}} : Spec_{\hat{O}}(\mathcal{H}_{SB}) \longrightarrow Spec_{\hat{O}}(\mathcal{H}_{MB})\,,
\end{equation*}
which, given an observable $\hat{O}$, maps linearly from the spectrum of $\hat{O}$ acting on the single-body Hilbert-Space to the spectrum
of $\hat{O}$ acting on the many-body Hilbert-Space. 
In this section, we will show that this lifting map introduces a clear separation between properties that are universal to many-body systems
and system-specific quantities. Moreover, the study of the lifting map will expose the elementary modes through which single-body spectral
information is transferred into the many-body spectrum.
We will be focusing on the Hamiltonian operator, for the purpose of computing the MBDoS, and as such we will consider a given single-body
energy spectrum as our only input for system-specific quantities.

In practice we introduce a matrix representation, denoted $F$ for convenience, of the Hilbert-space lifting map $F_{\hat H}$ for the Hamiltonian
operator. $F$ contains the universal combinatorial information : its matrix elements are occupation numbers $n_k$ for the $k$-th single-body
level, with $0 \le k \le L-1$ where $L$ is the number of single-body states.
Each row contains a valid many-body configuration containing $N = \sum_{k=0}^{L-1} n_k$ particles with at most $R$ allowed particles per
single-body state, i.e. $\forall k, n_k \le R$. $R$, called the restriction, is used to distinguish particle species : $R=N$ for bosons,
$R=1$ for fermion, or any particular value suitable to account for additional quantum numbers in the single-body picture if needed.
The matrix representation $F$ has $L$ columns and $C_R(L,N)$ rows which represent the number of many-body configurations : $C_1(L,N) = C_L^N$
for fermions and $C_N(L,N) = C_{L-1+N}^{L-1}$ for bosons. $F$ maps linearly between the representation spaces :
\begin{equation*}
    F : \mathbb{R}^L \longrightarrow \mathbb{R}^{C_R(L,N)} \,,
\end{equation*}
where the macroscopic parameters ($L$,$N$,$R$) fully characterize the universal structure of any identical particle system.

$F$ acts on the system-specific single-body spectrum $\epsilon = (\epsilon_0,\dots,\epsilon_{L-1})$ as follows :
\begin{align*}
    E &= F \; \epsilon \,,\\
    E(n) &= \sum_{k=0}^{L-1} n_k \; \epsilon_k \,,
\end{align*}
where $E$ denotes the vector of many-body energy values and $E(n)$ denotes a particular many-body energy value for the many-body configuration
denoted $n = (n_0,..,n_{L-1})$. The above equations implement the separation between universal information ($F$) and system-specific quantities
($\epsilon$).

Importantly, the combinatorial structure encoded in $F$ contains substantial redundancy in occupancy levels (i.e. columns of $F$) caused by the
full enumeration of many-body configurations. We therefore expect the lifting map to admit a compressed representation in which information is
organized into a hierarchy of symmetry-induced components.
More precisely, this redundancy manifests itself both with a very large kernel ($\dim \ker F = C_R(L,N) - L$) and highly degenerate non-zero
singular values which we can expose by performing the singular value decomposition (SVD) of $F$. We write :
\begin{equation*}
    F = \sum_{\ell = 0}^{L-1} \Sigma_\ell U^L_\ell \otimes (V^L_\ell)^\dagger \,,
\end{equation*}
where $\otimes$ is the outer product, $\Sigma_\ell$ are the singular values and $U^L_\ell$ / $V^L_\ell$ are the left / right singular vectors,
respectively.

In particular, the right-singular vectors $V^L_\ell$ can be obtained explicitly by diagonalizing the $F^\dagger \: F$ matrix which is a
$L\times L$ circulant matrix with only two distinct values, revealing the redundancy of information in $F$. Its eigenvectors are known and
involve discrete Fourier modes :
\begin{equation*}
    V_\ell^L = \frac{1}{\sqrt{L}} \left( 1, \omega_L^{\ell},\omega_L^{2 \ell}, \dots, \omega_L^{(L-1)\ell} \right) \,,
\end{equation*}
where $\omega_L = \exp(2\pi i / L)$ denotes the $L$-th root of unity. Note that, as expected, it can be shown that singular values $\Sigma_\ell$ only
take two distinct values \cite{lefevre2023many}.

On the other hand, the left-singular vectors $U^L_\ell$ cannot be accessed directly and need to be obtained using the relation :
\begin{equation*}
    F \: V^L_\ell = \Sigma_\ell U^L_\ell \,,
\end{equation*}
which requires to use the $F$ matrix directly. In order to avoid this, we will instead use the fact that the ordering of many-body configurations
within the $F$ matrix is arbitrary : we can disregard the ordering of the $U^L_\ell$ vector components and interpret them as a distribution of
complex values. For a given many-body configuration $n$ the associated complex value $U^L_\ell(n)$ is written :
\begin{equation}
    U^L_\ell(n) = \sum_{k=0}^{L-1} n_k \omega_L^{k\ell} \,,
    \label{eq:Ul}
\end{equation}
where we absorbed the singular value $\Sigma_\ell$ and the normalization $1/\sqrt{L}$ into the definitions of $U^L_\ell(n)$ and $V^L_\ell$
respectively to simplify notations from now on. We observe that $U^L_\ell(n)$ is the discrete Fourier transform of the occupation numbers $n_k$, a
direct manifestation of the periodicity of the columns of $F$. The set of $U^L_\ell$ distributions inherits the universal characterization of the
many-body system from $F$ and only depends on the macroscopic variables $L$, $N$ and $R$.

Finally, we can use the SVD of $F$ to perform a resummation of the many-body energy spectrum :
\begin{equation}
    E = F \: \epsilon = \sum_{\ell=0}^{L-1} U^L_\ell \: \tilde{\epsilon}^L_\ell \,,
    \label{resum}
\end{equation}
where $\tilde{\epsilon}^L_\ell = (V^L_\ell)^\dagger \cdot \epsilon$ are the Fourier coefficients of the single-body spectrum. This equation recasts
the complete separation of information into the set of universal distributions $U^L_\ell$ encoding the combinatorial structure of the lifting map,
and the effective coefficients $\tilde{\epsilon}^L_\ell$ containing all system-specific information. The many-body spectrum is therefore represented
as a superposition of universal modes weighted by system-dependent amplitudes : those are the core objects we will study from now on.

\begin{figure}
	\centering
	\includegraphics[width=0.8\linewidth, trim={.1cm .1cm 0 0}, clip]{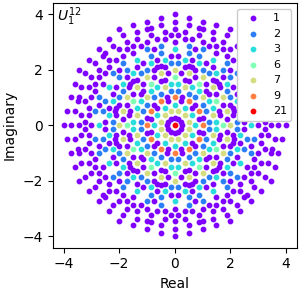}
	\caption{
		\textbf{Degeneracies in the distribution of the $U^L_\ell$ components in the complex plane for $L=12$,
        $N=4$ and $\ell = 1$ in the bosonic case.}
        The color of each point represents the degeneracy at their respective $U^L_\ell(n)$ values with
        exact counts given in the legend.
	}
	\label{fig:UDist-Degeneracies-Boson-L12N4l1}
\end{figure}

\subsection{Renormalization Structure of the Lifting Map}
\label{sec:q-sec}

Focusing on the set of universal $U^L_\ell$ distributions, our goal here is to characterize them and partition them into symmetry-induced sectors.
$U^L_\ell$ distributions are valued in the complex plane (\cref{eq:Ul}), mapping many-body configurations to complex values. However, multiple
configurations can map to the same complex value, this is depicted in \cref{fig:UDist-Degeneracies-Boson-L12N4l1} where the degeneracies of each
point on the complex plane are color-coded, we refer to these as degeneracy classes.

The emergence of degeneracy classes indicates that the lifting map partitions many-body configurations into equivalence classes carrying common spectral
information. This suggests an interesting classification of the combinatorial information directly induced by the lifting map which is exactly what we
pursue : an emerging intrinsic structure rather than a manually imposed hierarchy. To better understand this classification, we look at symmetries that
preserve $U^L_\ell$ distributions as a whole : these can be represented by the action of a symmetry group on the complex values rather than individual
configurations.

Investigating a single $U^L_\ell$ distribution, we observe a rotational symmetry (with angle $2\pi/L$) which preserves degeneracy classes : this is a
direct consequence of the set of $U^L_\ell(n)$ values across all configurations being invariant by circular permutations of the roots of unity in
\cref{eq:Ul}. This gives a first insight on the link between geometric and combinatorial properties (see \cite{lefevre2023many}).

However, to identify the maximal set of transformations preserving the information content of the lifting map, we seek symmetries acting simultaneously
on all $U^L_\ell$ distributions across $\ell$-modes. The maximal symmetry group acting on the set of $U^L_\ell$ distributions is the Galois group of the
cyclotomic field $\mathbb{Q}(\omega_L)$ where $U^L_\ell$ distributions take their values \cite{lefevre2026symmetries}. This Galois group can be represented
with Froebenius morphisms of the form :
\begin{equation*}
    \sigma^L_k( \omega_L^p ) = \omega_L^{kp} \,,
\end{equation*}
which act \emph{linearly} on roots of unity in $\mathbb{Q}(\omega_L)$. The Galois group itself is given by the automorphisms :
\begin{equation*}
    G_L = \{ \sigma^L_k \: | \: k \in \{0,\dots,L-1\}, \gcd(L,k) = 1 \} \,.
\end{equation*}
This Galois group acts on the set of $U^L_\ell$ distributions and preserves degeneracy classes (this is a consequence of the Froebenius automorphisms
inducing a permutation representation of $G_L$ acting on the set of $U^L_\ell(n)$ values). More generally, this Galois group defines the equivalence
classes of $U^L_\ell$ distributions and we call the associated symmetry $\ell$-symmetry (see \cite{lefevre2023many,lefevre2026symmetries}).
In particular, each $\ell$-symmetry equivalence class contains a distribution $U^L_d$ such that $d$ divides $L$. We will denote $q=L/d$ and refer to
these equivalence classes as $q$-sectors. As we will show in this section, $q$-sectors correspond to distinct resolutions of the lifting map, with
larger $q$ retaining finer information about many-body configurations and smaller $q$ providing progressively coarser descriptions.

Moreover, the full set of Froebenius morphisms form a semi-group which can be understood as a renormalization group. Each $q$-sector has its own Galois
group $G_q$ which acts on the corresponding $U^L_\ell$ distributions within. Froebenius morphisms which do not belong to $G_q$ can still act on the
$U^L_\ell$ distributions of this $q$-sector, but will instead map to another $U^L_\ell$ distribution which belongs to another $q$-sector : they provide
transitions between $q$-sectors and induce a sector flow.

\begin{figure*}
    \centering
    \includegraphics[width=\linewidth
    ]{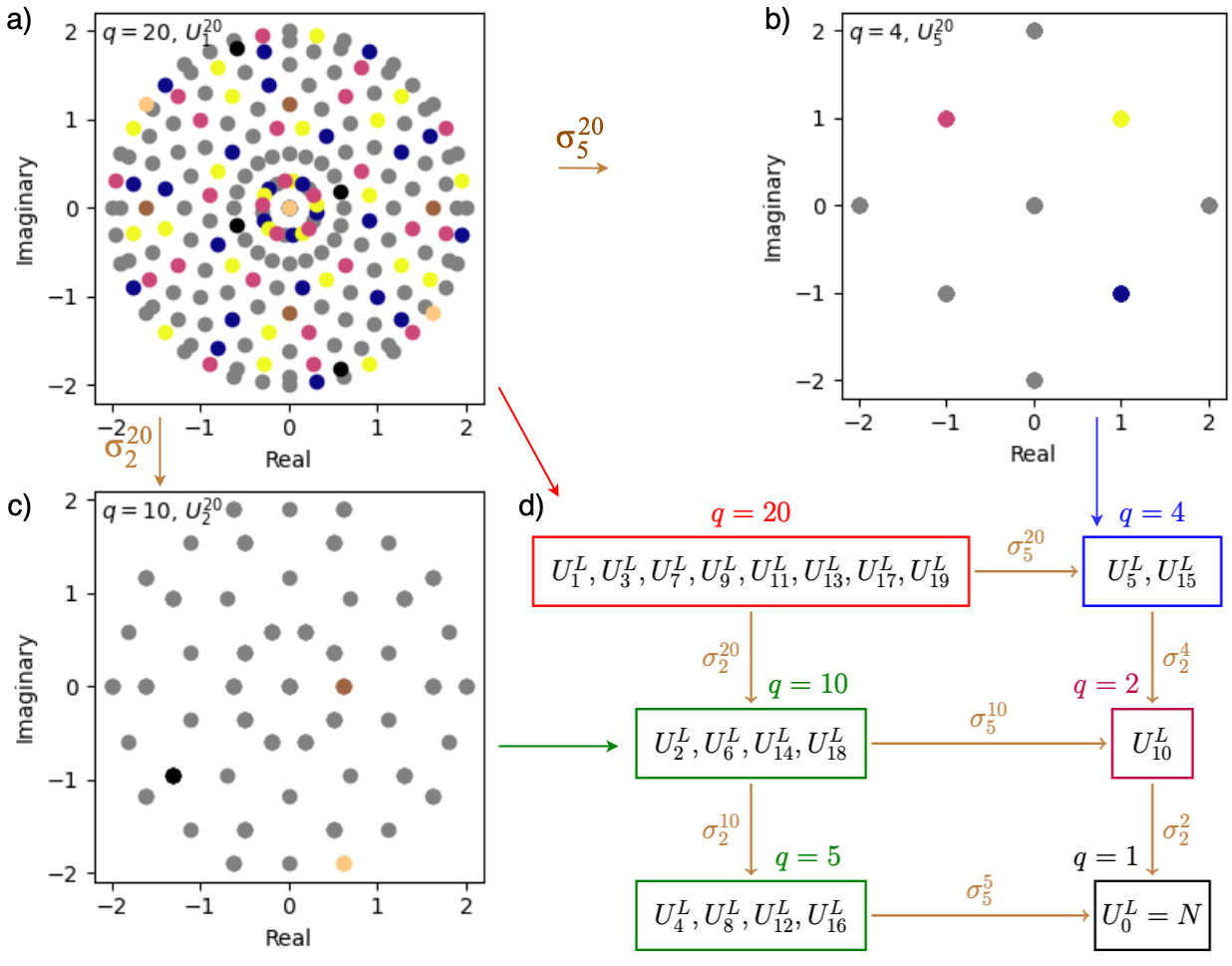}
    \caption{
        \textbf{Sector flow diagram for the lifting map with $L=20$ single-body levels.}
        We show the $L=20$ sector flow diagram (d) and detail how the $q$-sector transitions from the $q=20$ sector to both the $q=10$ and $q=4$ sectors
        occur under the action of the Froebenius morphisms.
        Diagrams for the $U^L_\ell$ distributions with $N=2$ corresponding to those sectors are shown with color-coding to indicate how degeneracy classes
        from $U^{20}_1$ (a) are merged under the action of $\sigma^q_2$ to yield $U^{20}_2$ (c) and the action of $\sigma^q_5$ to yield $U^{20}_5$ (b).
    }
    \label{fig:sectorflow}
\end{figure*}

We can represent this flow with a sector flow diagram as shown in \cref{fig:sectorflow} (d), where one can follow the flow from the red top-left $q=L=20$
sector to the black bottom-right $q=1$ "sink" sector. Brown arrows indicate which Froebenius morphism performs which $q$-sector transition.
Sector flow diagrams commute : the sector flow is path-independent \cite{lefevre2026symmetries}. It follows that the sector flow can be seen as a
renormalization flow where the Galois group becomes larger along the flow until it contains all Froebenius (auto)morphisms in the last sink sector.
\cref{fig:sectorflow} (a,b,c) show representative $U^L_\ell$ distributions for the $q=L=20$(red), $q=4$(blue) and $q=10$(green) where $U^L_\ell(n)$ values
in the complex plane are color-coded to illustrate how degeneracy classes are merged between $q$-sectors under the action of Froebenius morphisms outside
the Galois group.

We showed the sector flow induces a progressive merging of degeneracy classes : as a result, information about the many-body configuration space becomes
increasingly compressed along the flow, as one can see from \cref{fig:sectorflow} (a,b,c) where $U^L_\ell$ distributions become coarser. $q$-sectors
therefore define a hierarchy of resolutions of the lifting map, ranging from fine-grained descriptions that distinguish many-body configurations in
detail to coarse-grained descriptions that retain only global spectral information. This hierarchy is universal : it depends only on the combinatorial
structure of the many-body Hilbert space through $L$, and is independent of the particular single-body spectrum. More precisely, the complete sector flow
diagram is fully determined by the prime decomposition of the macroscopic variable $L$.

\subsection{Symmetry-Induced Quantum Numbers}
\label{sec:Invar}

Having established that $q$-sectors provide a hierarchy of resolutions of the lifting map, we now seek a description of the information contained within
a given sector. Our goal is to identify the quantities that distinguish degeneracy classes and determine how this information is progressively lost along
the renormalization flow. We will show that each $q$-sector admits a set of invariant coordinates $\{I^q_k\}_{\varphi(q)}$, where $\varphi(q)$ is the Euler
totient function. The collection of these invariants across all $q$-sectors form a complete set of quantum numbers equivalent to the occupation-number
description of the many-body Hilbert space (see \cite{lefevre2026symmetries}).

We already introduced the cyclotomic field where $U^L_\ell(n)$ is valued : however, looking back at \cref{eq:Ul}, $L$-roots of unity are not linearly
independent. As a vector space over $\mathbb{Q}$, the cyclotomic field $\mathbb{Q}(\omega_q)$ of a given $q$-sector has a basis for which there are several
choices (see \cite{lefevre2026symmetries}), for our particular purpose here we will use the simplest one given by :
\begin{align*}
    \mathcal{B}_q &= \{ 1, \omega_q, \omega_q^2, \dots, \omega_q^{\varphi(q)-1} \}\,, \\
    V^q_1 &= T_q \cdot \mathcal{B}_q \,,
\end{align*}
where we introduce the $q \times \varphi(q)$ rectangular matrix $T_q$ which maps back to the full set of $q$-roots of unity in this $q$-sector
(see derivation in \cref{appendix:tq}).

This cyclotomic basis introduces unique coordinates corresponding to each complex value $U^L_\ell(n)$, and therefore transforms the geometric structure of
the $U^L_\ell$ distributions into a discrete set of labels for each degeneracy class : in other words, these coordinates leave the complex value invariant
across many-body configurations within the degeneracy class. We call these coefficients invariants and denote them $I^q_k$, they yield a linear
expansion in the $\mathcal{B}_q$ basis of a given $q$-sector :
\begin{equation}
    U^q_1(n) = \sum_{k=0}^{\phi(q)-1} I^q_k \omega_q^{k} \,,
    \label{eq:UlI}
\end{equation}
where we chose the $U^q_1 \sim U^L_\ell$ distribution as the representative to express complex values with $q$-roots of unity in the corresponding
$q$-sector : we simply rewrite $\omega_L^{k\ell} = \omega_q^k$ since $\ell$ divides $L$ (refer to the folding interpretation of $q$-sector transitions
in \cite{lefevre2023many} for further details).

Remarkably, the collection of invariants across all $q$-sectors is complete : no information about the many-body configuration is lost. These invariants
therefore provide an alternative description of many-body Hilbert space in terms of \emph{symmetry-induced quantum numbers}. They form a hierarchy of
quantum numbers along the renormalization flow :
\begin{equation*}
    \{I^{q=L}_k\} \rightarrow \{I^{q_1}_k\} \rightarrow \{I^{q_2}_k\} \rightarrow \dots \,.
\end{equation*}
One can observe that sector-flow transitions correspond to systematic reductions in the number of independent quantum numbers, since the dimension of the
cyclotomic field $\mathbb{Q}(\omega_q)$ becomes smaller when $q$ is smaller : this provides an explicit information-hierarchy characterization of the
renormalization structure.

\section{Symmetry-Induced Generating Function}
\label{sec:GF}

To understand the information that is discarded under the aforementioned renormalization flow, we need a precise counting of the size of each degeneracy
class of the lifting map. As we will see, the flow of the symmetry-induced quantum numbers $I^q_k$ along $q$-sectors allows us to solve this counting
problem in an efficient manner.

We start by introducing the generating function for identical particle systems using a single variable polynomial :
\begin{equation}
    g(X) = \prod_{k=0}^{L-1} \left[ \sum_{n_k=0}^{R} X^{n_k} \right] = \sum_{p=0}^{LR} C_R(L,p) X^p \,,
    \label{eq:basegenfunc}
\end{equation}
which counts the total number of many-body configurations in the coefficient of $X^N$.

We can use the $T_q$ matrix we introduced in the previous section to construct the generating function of degeneracy classes. Recall that the $T_q$ matrix
maps from the cyclotomic basis of a given $q$-sector back to the full set of $q$-roots of unity (as shown in \cref{appendix:tq}), it can also be used to quantify the
value of a quantum number $I^q_k$ with respect to an occupancy number $n_k$ using the relations :
\begin{equation}
    \begin{split}
        I^q_{k'}(n_k) &= n_k \cdot (T_q)_{k\%q,k'}\,, \\
        I^q_{k'} &= \sum_{k=0}^{L-1} I^q_{k'}(n_k) \,,
    \end{split}
    \label{eq:deltaiq}
\end{equation}
where $I^q_{k'}(n_k)$ is the $k$-th single-body level contribution to the value of $I^q_{k'}$.

Combining the above expressions \eqref{eq:basegenfunc} and \eqref{eq:deltaiq}, we promote the generating function to a multi-variable polynomial where $X$
is still encoding the constraint on the total number of particles $N$, while we introduce a set of auxiliary variables $\{Y_{k'}\}_{\varphi(q)}$ conjugate
to the quantum numbers $I^q_{k'}$ which allow the generating function to resolve individual degeneracy classes within a given $q$-sector. We write :
\begin{equation}
    \begin{split}
        G_q(X,\{Y_{k'}\})
        &= \prod_{k=0}^{L-1} \left[
            \sum_{n_k=0}^R \left(
                    X \prod_{k'=0}^{\phi(q)-1} Y_{k'}^{(T_q)_{k,k'}}
                \right)^{n_k}
            \right]\,, \\
        &= \prod_{k=0}^{L-1} \left[
            \sum_{n_k=0}^R X^{n_k} \left(
                \prod_{k'=0}^{\varphi(q)-1} Y_{k'}^{I_{k'}^q(n_k)}
            \right)
        \right]\,,
    \end{split}
    \label{eq:modgenfunc}
\end{equation}
where the size of a degeneracy class in a $q$-sector given by a particle count $N$ and a set of quantum number values $I^q_k$ is contained in the coefficient
of the $X^N \prod_k Y_k^{I^q_k}$ term.

Following the same strategy, we can combine all the $q$-sectors together into a single generating function where the set of auxiliary variables $\{Y_{q,k'}\}_L$
now covers all $q$-sectors at once :
\begin{equation}
    \begin{split}
        G(X,\{Y_{q,k'}\})&=\prod_{k=0}^{L-1}\left[\sum_{n_k=0}^R X^{n_k}\prod_{q|L}\left[\prod_{k'=0}^{\phi(q)-1}Y_{q,k'}^{I^q_{k'}(n_k)}\right]\right] \,, \\
        &= \sum_{p,\{I^q_{k'}\}_{q,k'}} c_{ \{ p, \{I^q_{k'}\}_{q,k'} \} } X^p \prod_{q,k'} Y_{q,k'}^{I^q_{k'}} \,,
    \end{split}
    \label{eq:gprime}
\end{equation}
where the second line shows how the polynomial is expanded such that each coefficient $c_{p,\{I^q_{k'}\}_{q,k'}}$ corresponds to the size of the degeneracy class
associated with $p$ particles and quantum number values $\{I^q_{k'}\}_{q,k'}$. One can find detailed examples of calculations using this generating function in
\cref{appendix:gf}.

Using the above tools, we can now give a qualitative characterization of how information is lost along the sector flow diagram. This immediately leads us to a
truncation strategy by selectively removing $q$-sectors from the generating function \eqref{eq:gprime}. Looking at the paths in a given sector flow diagram, we
can systematically choose a subset of $q$-sectors to remove based on their hierarchical position in the renormalization flow. Truncating selected $q$-sectors
corresponds to discarding specific layers of symmetry-induced quantum numbers : the sector hierarchy therefore provides a controlled coarse-graining of the
many-body Hilbert space, enabling systematic tradeoffs between spectral resolution and computational complexity.

\section{Hierarchical Observable Reconstruction}

Having identified a hierarchy of symmetry-induced quantum numbers, we now show how this hierarchy allows reconstruction of physical observables and enables
controlled coarse-graining of the many-body Hilbert space. Taking the Hamiltonian operator as an example, we apply the renormalization flow to expand the modes
of the lifting map $U^L_\ell$ in \cref{resum} which allows us to expand the many-body energy defined by a set of quantum number values $\{I^q_{k'}\}_{q,k'}$ over
the sector flow diagram and we obtain (see \cref{appendix:e}) :
\begin{equation}
    E_{\{I^q_{k'}\}_{q,k'}} = N \tilde{\epsilon}^L_0 +
        \sum_{\underset{\ell|L,\ell<L}{q=L/\ell}} \:
        \sum_{\sigma^q_p \in G_q} \:
            \underbrace{
                \sum_{k'=0}^{\phi(q)-1}I^q_{k'}\sigma^q_p(\omega_q^{k'})
            }_{U^q_p}
            \cdot \tilde\epsilon^q_p
    \,,
    \label{eq:resum}
\end{equation}
where the first term for the sink sector is set apart to avoid notational issues, the outer sum over the valid values of $q$ (when $\ell$ divides $L$) spans
all $q$-sectors, the next sum uses the Galois group $G_q$ to span the $U^q_p$ distributions in the parent $q$-sector and the innermost sum iterates over the
symmetry-induced quantum numbers $I^q_{k'}$ to span the set of degeneracy classes in the parent distribution.

Equation \eqref{eq:resum} expresses the many-body energy as a hierarchical expansion over the renormalization structure of the lifting map. Each $q$-sector
contributes an independent symmetry-induced component weighted by the corresponding Fourier coefficient of the single-body spectrum. Truncating the expansion
at a chosen level of the hierarchy corresponds to removing entire families of symmetry-induced quantum numbers.

The hierarchical decomposition provides direct control over the spectral information affected by a truncation. Since the discarded contribution is entirely
encoded in the Fourier amplitudes $\tilde{\epsilon}^q_p$, the approximation error can be reduced by minimizing their magnitude through a suitable ordering of
the single-body spectrum. This procedure generalizes immediately to any other observable and single-body spectrum quantities, as long as the observable's
spectrum is discrete.

\section{Numerical Results}

\subsection{Symmetry-Induced MBDoS Computation}

\begin{figure*}[ht!]
    \centering
    \includegraphics[width=\linewidth]{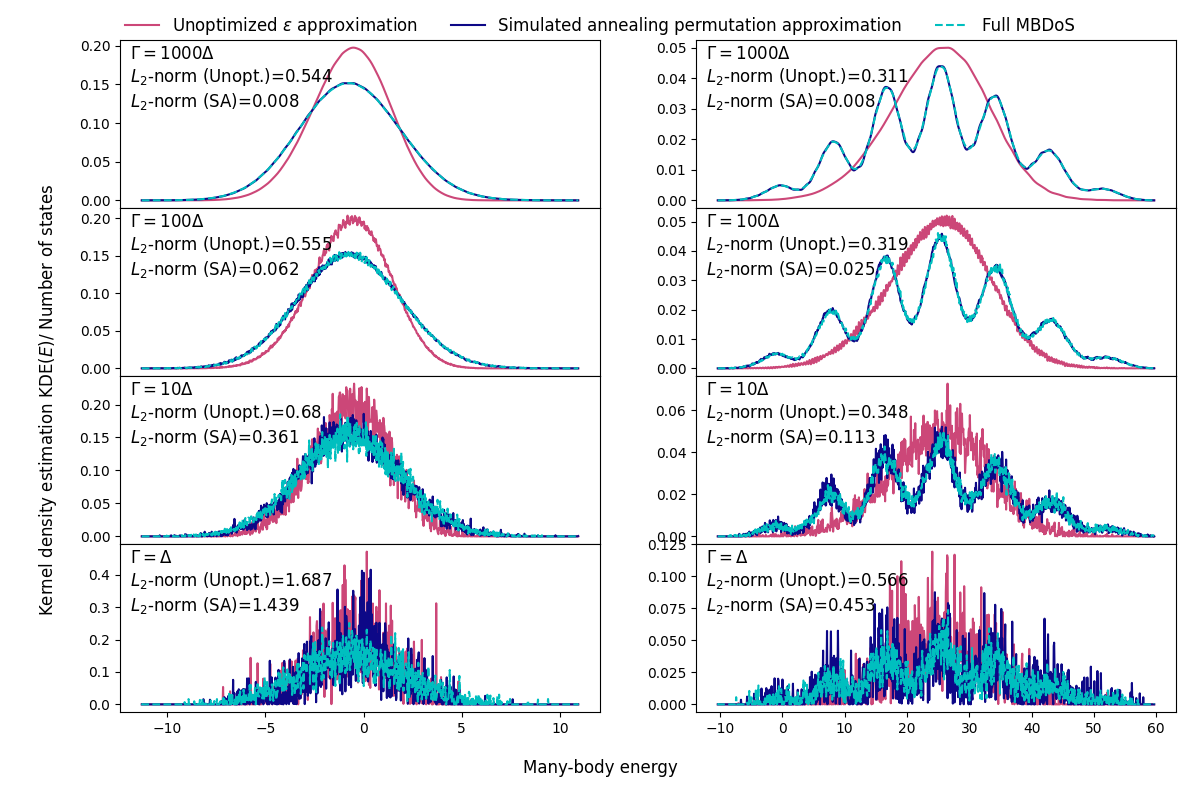}
    \caption{
        \textbf{Comparisons of MBDoS approximations obtained from hierarchical decomposition of the lifting map in the Gaussian and non-Gaussian cases.}
        We show normalized kernel density estimations (KDEs) of both the full MBDoS and approximations of the MBDoS obtained by discarding the $q=20$ sector
        for Gaussian (left column) and non-Gaussian (right column) single-body energy spectra.
        KDEs obtained from the full MBDoS are shown in dotted cyan lines, whereas the approximations obtained from the unoptimized labeling of single-body
        energies, and the labeling obtained from optimized simulated annealing are illustrated in pink and blue lines respectively.
        The fidelity between the approximated MBDoS and the exact MBDoS are evaluated using $L_2$-norms between the KDE curves.
        The KDEs are calculated from convolutions of the full MBDoS (or its approximations) with Gaussian kernel widths $\Gamma$ of varying orders of magnitudes
        of the mean level spacing $\Delta$. These values of $\Gamma$ provide different levels of energy resolutions when evaluating the MBDoS.
        As values of $\Gamma$ decrease, the $L_2$-norm values increase, alluding to lower approximation accuracies. This is due to the granularity of the MBDoS
        and its approximations ; for lower values of $\Gamma$, this granularity persists after convoluting with narrower kernels, as exposed by the sharp peaks
        in the KDEs.
    }
    \label{fig:MergedKDE}
\end{figure*}

In the context of the computation of the MBDoS, the group of permutations over the set $\{\epsilon_k\}_L$ of single-body energies is the symmetric group $S_L$
which contains $L!$ elements. This group being very large, we choose an oriented search approach based on a heuristic function which estimates how well the set
of $\tilde{\epsilon}^q_p$ values from a given truncated $q$-sector is minimized.

Simulated annealing is a good example of such an algorithm since the number of transpositions between two permutations (i.e. the branching path) is of the order
of $L$, much smaller than the size of $S_L$. Since truncation errors originate exclusively from the discarded Fourier amplitudes, a natural measure of the
importance of a $q$-sector is the total spectral weight carried by these amplitudes. This measure is typically implemented by the choice of a heuristic function
and is of utmost importance in the performance of such an algorithm. The numerical computations below will use :
\begin{equation*}
    h_q(\epsilon) = \frac{
            \frac{1}{L} \sum_{\mathrm{gcd}(q,p)=1} |\tilde\epsilon^q_p|^2
        }{
            \sum_{k=0}^{L-1} |\epsilon_k|^2
        } \,,
\end{equation*}
which is a second-order sum of the amplitudes in the discarded $q$-sector derived from Parseval's identity (calculation shown in \cref{appendix:lowerbound}). It quantifies the spectral weight carried by that
sector and therefore provides a proxy for its contribution to observable reconstruction. An estimate of the lower bound to achieve to obtain a good permutation
is given by :
\begin{equation}
    \min[h_q(\epsilon)] \sim \frac{
            \phi(q)^2 \times \Delta^2_{\mathrm{SBE}}
        }{
            L\sum_{k=0}^{L-1} |\epsilon_k|^2
        } \,,
    \label{eq:heuristicfunc}
\end{equation}
which can be used as a stopping criteria for the simulated annealing algorithm (this result is derived in \cref{appendix:lowerbound}).

To demonstrate this technique in practice, we choose a sample system of identical bosons with $L=20$ single-body levels and $N=6$ particles where the first
sector ($q=L=20$) in the flow diagram has been truncated. This sector was selected because it corresponds to the finest level of the hierarchy and therefore
contains the highest-resolution $U^L_\ell$ distributions while also generating the largest computational cost.

Since discrete MBDoS histograms does not accommodate comparisons, we instead evaluate coarse-grained spectral densities obtained through kernel density
estimation (KDE), which perform a convolution of the MBDoS against a broadening kernel with freedom in the choice of its width (a Gaussian kernel is enough
for our purpose here).
We obtain coarse-grained curves that can be compared using their $L_2$-norm over the full width of each spectrum. The freedom in the kernel width can be
used to account for different resolution targets on the MBDoS.

In \cref{fig:MergedKDE}, we chose several kernel widths $\Gamma$ in a range of values scaled from the fully enumerated many-body spectrum mean level spacing
$\Delta$. On the top row, the broadest resolution ($\Gamma = 1000\Delta$) probes the coarsest physically relevant scale in the many-body spectrum, given
by the gap between the many-body ground state and the first many-body excited state. On the bottom row, the finest resolution ($\Gamma = \Delta$) corresponds
to the smallest characteristic energy scale, given by the mean level spacing of the many-body spectrum.

We also explore two different choices for the single-body energy spectrum, which lead to a Gaussian MBDoS (on the left column) and a non-Gaussian MBDoS
(on the right column).

Fig. \ref{fig:MergedKDE} compares the numerical results of two choices of single-body spectrum orderings against the reference MBDoS computed from full
enumeration and shown with a dashed cyan curve. The pink curve represents the MBDoS computation performed using our framework with sector truncation but
without any further optimization, i.e. the single-body spectrum is left in increasing energy order. On the other hand, the dark blue curve shows the same
computation, with the re-ordering of the single-body spectrum implemented using simulated annealing with the heuristic function $h_q$ from above.

From top to bottom, we observe how the optimized ordering provides a significant qualitative improvement to the approximation of the true MBDoS, over all
considered resolution constraints. The same pattern is observed on the right in the non-Gaussian case where the optimized ordering is able to capture the
multiple extrema in the MBDoS. This demonstrates that controlling the spectral weights carried by discarded $q$-sectors enables selective preservation of
physically relevant information, allowing coarse-graining without destroying qualitative spectral structure.

\subsection{Recovering the Bose-Einstein Distribution}

Having reconstructed the MBDoS through the hierarchical decomposition of the lifting map, we can recover thermodynamic observables that depend only on the
spectral structure of the many-body Hilbert space. Here we provide a simple example of recovering the occupation number expectation value using :
\begin{equation}
    \begin{split}
        \braket{n_k}&=\frac{\sum_{n=0}^{N}n\,\mathrm{KDE}(E_i-n\epsilon_k,L-1,N-n)}{\mathrm{KDE}(E_i,L,N)}\,,\\
        &=\frac{\sum_{n=0}^{N}n\,\mathrm{KDE}(E_i-n\epsilon_k,L-1,N-n)}{\sum_{n=0}^N\mathrm{KDE}(E_i-n\epsilon_k,L-1,N-n)}\,,
    \end{split}
    \label{eq:nk}
\end{equation}
which follows from counting the number of many-body configurations compatible with a fixed occupation of the $k$-th single-body energy level.

\begin{figure*}
    \centering
    \includegraphics[width=\linewidth]{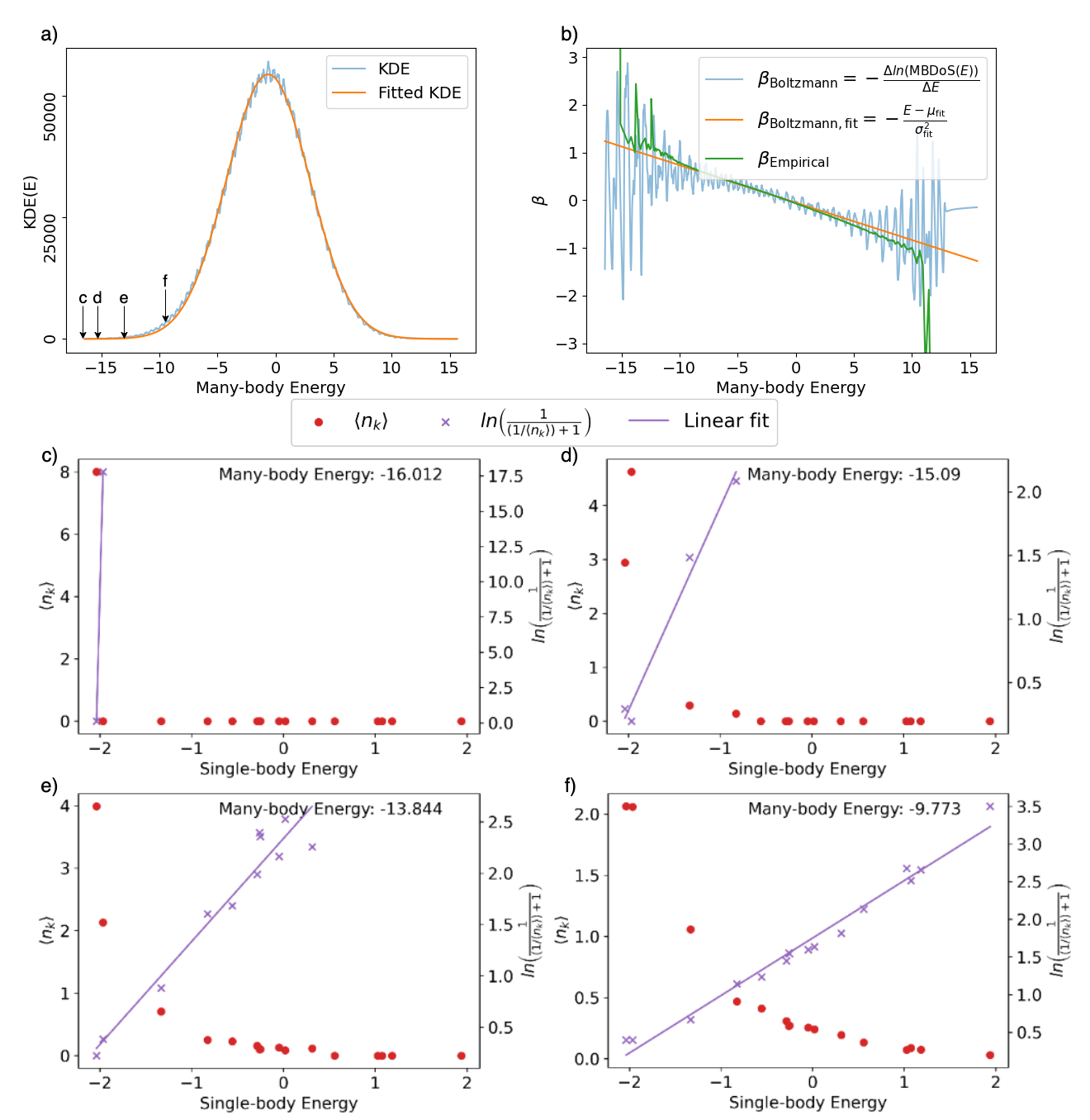}
    \caption{
        \textbf{Recovery of Bose-Einstein-like distributions for a system of $N=8$ identical bosons with $L=16$ single-body energy levels.}
        a) The many-body density of states (MBDoS) as a function of many-body energy $E$, obtained via a kernel density estimation (KDE) with kernel width
        $\Gamma=1000\Delta$ where $\Delta$ is the mean level spacing (blue), together with a Gaussian fit (orange).
        The MBDoS is globally well described by a Gaussian over the bulk of the spectrum, consistent with central-limit behaviour.
        Deviations appear only in the extreme low- and high-energy tails, where the MBDoS becomes sparse and finite-size effects dominate.
        The arrows c–f indicate the energies at which specific single-particle occupation analysis is displayed in panels c)–f).
        b) Microcanonical inverse temperatures $\beta(E)$ obtained from three independent procedures :
        (i) $\beta_{\mathrm{Boltzmann}}$ computed from the logarithmic derivative of the smoothed MBDoS (blue);
        (ii) $\beta_{\mathrm{Boltzmann,fit}}$, derived analytically from the Gaussian fit in a) (orange), which yields a strictly linear function of energy;
        (iii) $\beta_{\mathrm{Empirical}}$, extracted from linear fits to transformed single-particle occupancies (green).
        c)–f) Single-particle occupation numbers $\braket{n_k}$ at representative many-body energies indicated in a).
        The red points show $\braket{n_k}$ versus single-body energies $\epsilon_k$. Purple crosses represent the transformed quantity $\ln(\frac{1}{1/\braket{n_k}+1})$,
        which should be linear in $\epsilon_k$ for an effective Bose–Einstein distribution, with slope equal to the inverse temperature $\beta_{\mathrm{Empirical}}$.
        The purple line denotes the linear fit used to extract this slope.
        We observe significant differences between the bulk and edge cases in b). The rapid oscillations visible in $\beta_{\mathrm{Boltzmann}}$ occur
        predominantly near the spectral edges and arise from differentiation of a finite and discretely sampled density of states; numerical differentiation
        amplifies residual fluctuations of the KDE and discreteness. The deterioration of $\beta_{\mathrm{Emprirical}}$ linearity near the edges reflects
        breakdown of the canonical approximation.
    }
    \label{fig:placeholder}
\end{figure*}

We approximate this expectation occupancy by substituting MBDoS approximations where the largest $q$-sectors are truncated in the evaluations of the KDE of
the reduced MBDoS : KDE$(E_i-n\epsilon_k,L-1,N-n)$ for a $L=16, N=8$ system (see \cref{fig:placeholder}).
In subfigure a), we show the truncated sector MBDoS approximation depicted via a KDE (blue) with $\Gamma=1000\Delta$. We also plot a Gaussian fit (orange)
as a reference for the inverse temperature $\beta$ plots in subfigure b).
In subfigure b), we compare between three different estimates of the inverse temperature $\beta(E)$. The blue curve corresponds to the microcanonical
definition :
\begin{equation*}
    \beta_{\mathrm{Boltzmann}}(E) = \frac{\partial}{\partial E}\ln \mathrm{MBDoS}(E)\,,
\end{equation*}
computed numerically from the logarithmic derivative of the smoothed MBDoS. The orange curve shows $\beta_{\mathrm{Boltzmann,fit}}$, obtained analytically
from the Gaussian fit of subfigure a), which leads to a linear dependence of $\beta$ on $E$. The green curve shows $\beta_{\mathrm{Empirical}}$, extracted
from linear fits to transformed occupation numbers of the form $\ln\!\left(\frac{1}{\braket{n_k}}+1\right)$ as discussed below.

Subfigures c) to f) show the expectation values of the single-particle occupation numbers $\braket{n_k}$ at the energies indicated in subfigure a). The red
markers show $\braket{n_k}$ plotted as a function of the single-particle energies $\epsilon_k$. The purple crosses correspond to the transformed quantity
$\ln\!\left(\frac{1}{\braket{n_k}}+1\right)$, which should depend linearly on $\epsilon_k$ if the occupations follow a Bose-Einstein-like distribution with an
effective inverse temperature $\beta$. The purple lines indicate the linear fits used to extract the empirical inverse temperature
$\beta_{\mathrm{Empirical}}$ from the slope. That is to say that the Bose-Einstein-like occupation profile is recovered from the MBDoS even after $q$-sector
truncation, through counting microstates in \cref{eq:nk}.

At the lowest energy shown in subfigure c), the occupations are strongly localized in the lowest energy modes, producing a large positive effective inverse
temperature consistent with proximity to the ground-state spectral edge. At intermediate energies, subfigures d), e) and f), the transformed occupations
exhibit a clear linear dependence on $\epsilon_k$, indicating a well-defined effective temperature and approximate equivalence between the empirical and
Boltzmann descriptions in the bulk of the spectrum.

Overall, the figure demonstrates that the MBDoS is well approximated by a Gaussian over the bulk of the spectrum, implying a single smooth entropy branch
and an approximately quadratic entropy function $S(E)$. The resulting inverse temperature $\beta(E)$ is therefore analytic and nearly linear in the central
energy region. The pronounced variations observed near the spectral edges originate from finite-size and boundary effects rather than from thermodynamic
non-analyticities.

The present framework therefore provides a universal non-interacting reference MBDoS, while interactions can subsequently be incorporated through transition
probabilities between many-body configurations. In particular, we emphasize on the probability of transition being computed explicitly from methods such as
random interactions using the Dyson loop equation (\cite{ithier2018statistical}) in order to go beyond the equiprobability assumption from the microcanonical
ensemble. This provides a route toward studying non-ergodic many-body phenomena such as many-body localization
\cite{schreiber2015observation,choi2016exploring,smith2016many} and quantum scars \cite{bernien2017probing,turner2018weak,serbyn2021quantum}, where standard
assumptions of ergodic thermalization can fail.

\section{Conclusion}

In this paper, we introduced a framework based on the Hilbert-space lifting map, which describes how single-body information becomes many-body information.
We revealed its intrinsic symmetry, driven by Galois groups, which preserves degeneracy classes of many-body configurations in the spectrum of the lifting
map and thus encodes the combinatorial information of many-body systems.

The resulting symmetry-induced quantum numbers provide a hierarchical description of the many-body Hilbert space into equivalence classes we call
$q$-sectors. They can be represented in a sector flow diagram which defines a renormalization flow that progressively coarse-grains spectral information
while retaining a complete representation of the observable. This reveals that the number of single-body states $L$, usually considered a microscopic
parameter, acts as an emergent structural variable organizing many-body information.

Using these quantum numbers, we constructed a generating function that determines the cardinality of degeneracy classes and enables efficient computation of
the MBDoS. Applying the framework to bosonic systems, we demonstrated controlled truncation of the hierarchical structure and showed that optimizing the
distribution of spectral information across sectors significantly improves MBDoS reconstruction. We further showed that the resulting spectral information,
even after truncation, is sufficient to recover occupation statistics and Bose-Einstein-like behavior directly from the combinatorial structure.

The separation between universal Hilbert-space structure and system-dependent quantities allows for scalable computational strategies based on reusable
sector flow diagrams and generating functions. Moreover, the sector hierarchy suggests new coarse-graining strategies that are not imposed by spatial
locality but emerge from the information content of the Hilbert-space lifting map.

As recent experiments and quantum simulators increasingly require access to local spectral information in restricted energy windows, our framework could
be used to infer energy ranges and their overlaps, on a per-$q$-sector basis, without full diagonalization. This technique would allow for significant
optimizations based on symmetry-induced quantum numbers.

Although we focused on the Hamiltonian, the lifting-map construction applies to arbitrary observables with discrete spectra. Extending our framework to
multiple observables and their commutation relations may provide a systematic approach for uncovering universal structures underlying many-body quantum
systems.

\section{Author Contribution Statement}

H.L. and R.L. carried out the mathematical conceptualization, formulation and derivations.
H.L. and R.L. developed the physical interpretation and numerical applications.
H.L. performed the numerical data testing and the preparation of figures.
G.I. supervised the project.
All authors contributed to the writing of this article.

During the preparation of this manuscript, the authors used ChatGPT (OpenAI) to assist with language editing, improve clarity of exposition, and suggest
improvements to the organization of the manuscript. All scientific ideas, mathematical derivations, computational methods, figures, analyses, and
conclusions were developed, verified, and approved solely by the authors, who reviewed and edited all AI-assisted text and take full responsibility for
the content of the manuscript.

\FloatBarrier
\bibliography{bibliography.bib}

\clearpage
\newpage
\appendix

\section*{Appendix}
\addcontentsline{toc}{section}{Appendix}

\section{Transfer Matrix \texorpdfstring{$T_q$}{tq} Derivation}
\label{appendix:tq}

\subsection{The mapping of \texorpdfstring{$q$-sector basis}{}}
Here, we showcase a method of analytically evaluating the $q\times\varphi(q)$ transfer matrix $T_q$ that maps a $q$-sector basis $\mathcal{B}_q = \{ 1, \omega_q, \omega_q^2, \dots, \omega_q^{\varphi(q)-1} \}$ to the set of $q$-th roots of unity $V^q_\ell=( 1, \omega_L^{\ell},\omega_L^{2 \ell}, \dots, \omega_L^{(L-1)\ell})$. Explicitly, this relations is written as :
\begin{equation*}
\begin{split}
    V^q_1 &=T_q \cdot \mathcal{B}^q \,,\\
\end{split}
\label{eq:transfermatrix}
\end{equation*}
and the element-wise expansion of this matrix equation is :
\begin{equation}
    \omega_q^p = \sum_{k=0}^{\phi(q)-1} (T_q)_{p,k} \; \omega_q^k\,,
\label{eq:basisexpansion}
\end{equation}
where $p \in \{0,1,\dots,q-1\}$, and $(T_q)_{p,k}$ is the element of $T_q$ at row $p$ and column $k$.
Note that the first $\varphi(q)$ rows form an identity matrix from our choice of basis.

From this expansion, we derive an inductive method to expand successive powers of $\omega_q$. We first write two expressions :
\begin{align*}
    \omega_q^{p+1} &= \sum_{k=0}^{\phi(q)-1} (T_q)_{p+1,k} \; \omega_q^k \\
    \omega_q^{p+1} &= \sum_{k=0}^{\phi(q)-1} (T_q)_{p,k} \; \omega_q^{k+1}\,,
\end{align*}
where the first line is the expansion of $\omega_q^{p+1}$ and the second line is the expansion of $\omega_q^p$ multiplied on both side by $\omega_q$.
After solving and identifying coefficients in the $\mathcal{B}_L$ basis, we obtain the induction relations :
\begin{align*}
    (T_q)_{p+1,0} &= (T_q)_{p,\varphi(q)-1} \: (T_q)_{\varphi(q),0} \\
    (T_q)_{p+1,k} &= (T_q)_{p,k-1} + (T_q)_{p,\varphi(q)-1} \: (T_q)_{\varphi(q),k} \\
     & \forall k \in \{1,\dots,\varphi(q)-1\}\,,
\end{align*}
where we still need to compute the $(T_q)_{\varphi(q),k}$ coefficients to allow the induction to start : they are the coefficients for the expansion
of $\omega_q^{\varphi(q)}$.

These coefficients can be easily obtained by using the cyclotomic polynomial $\Phi_q(x)$ which satisfies $\Phi_q(\omega_q) = 0$, allowing us to find the
expansion of $\omega_q^{\varphi(q)}$ immediately by expressing the leading term in $\Phi_q(\omega_q)$ in terms of the lower degree monomials.
The induction relations can now be rewritten :
\begin{align*}
    (T_q)_{p+1,0} &= - \phi_0 \: (T_q)_{p,\varphi(q)-1} \\
    (T_q)_{p+1,k} &= (T_q)_{p,k-1} - \phi_k \: (T_q)_{p,\varphi(q)-1} \\
     & \forall k \in \{1,\dots,\varphi(q)-1\}\,,
\end{align*}
where $\phi_k$ is the coefficient of $x^k$ in the cyclotomic polynomial $\Phi_q(x)$ :
\begin{align*}
    \Phi_q(x) = x^{\varphi(q)} + \sum_{k=0}^{\varphi(q)-1} \phi_k x^k\,.
\end{align*}

\subsection{The mapping of new quantum numbers}

These $T_q$ matrices can be used to quantify the contribution of single-body occupancies $\{n_k\}_L$ to the new quantum numbers $\{I^q_p\}_q$ of
the corresponding $q$-sector. This can be shown from \cref{eq:Ul} by considering the mode $\ell$ which divides $L$, within the $q=L/\ell$ sector :
\begin{equation*}
    U^L_\ell(n) = \sum_{k=0}^{L-1} n_k \omega_L^{k\ell} = \sum_{k=0}^{L-1} n_k \omega_q^k = \sum_{k=0}^{L-1} n_k \omega_q^{k\%q} \,,
\end{equation*}
where in the last expression, we emphasize the periodicity of $q$-roots of unity explicitly : for $k\ge q$, $\omega_q^k$ can be reduced to
$\omega_q^{k\%q}$.

Using \cref{eq:UlI}, $U^L_\ell(n)$ can also be written as :
\begin{equation*}
    U^L_\ell(n) = \sum_{p=0}^{\varphi(q)-1} I^q_p \omega_q^p
        = \sum_{p=0}^{\varphi(q)-1} \left( \sum_{k=0}^{L-1} I^q_p(n_k) \right) \omega_q^p \,,
\end{equation*}
where $I^q_p(n_k)$ is the $k$-th single-body level contribution to the value of $I^q_p$.

From the above two expressions for $U^L_\ell(n)$, we deduce :
\begin{equation*}
    \sum_{k=0}^{L-1} n_k \omega_q^{k\%q} = \sum_{p=0}^{\varphi(q)-1} \left( \sum_{k=0}^{L-1} I^q_p(n_k) \right) \omega_q^p \,,
\end{equation*}
where we now use the $T_q$ matrix and \cref{eq:basisexpansion} to expand roots of unity on the left hand side. This yields :
\begin{equation*}
    \sum_{p=0}^{\varphi(q)-1} \left( \sum_{k=0}^{L-1} n_k (T_q)_{k\%q,p} \right) \omega_q^p
        = \sum_{p=0}^{\varphi(q)-1} \left( \sum_{k=0}^{L-1} I^q_p(n_k) \right) \omega_q^p \,.
\end{equation*}
where we re-ordered sums on the left hand side.

Using the fact that $\{\omega_q^p\}_{\varphi(q)}$ is a basis in the cyclotomic field of this $q$-sector, we can identify coefficients to obtain :
\begin{equation*}
    \sum_{k=0}^{L-1} n_k (T_q)_{k\%q,p} = \sum_{k=0}^{L-1} I^q_p(n_k) \,,
\end{equation*}
and since occupation numbers also form a basis in the configuration space, we have :
\begin{equation*}
    n_k(T_q)_{k\%q,p} = I^q_p(n_k) \,,
\end{equation*}
recovering \cref{eq:deltaiq}. 

\begin{widetext}
\section{Generating Function  \texorpdfstring{$G(X,\{Y_{q,k'}\})$}{gf} Examples}
\label{appendix:gf}

\subsection{Example : Fermions ($R = 1$), $L=6, N=2, \ell=1$}
Choosing small systems such that we can perform the full expansion of the generating function by hand,
in this example, we examine the $q= L/\ell = 6$ sector, for which the $T_6$ matrix can be calculated from the method outlined in \cref{appendix:tq} :
\begin{equation*}
    T_6^\top = \begin{bmatrix}
        1 & 0 & -1 & -1 &  0 &  1 \\
        0 & 1 &  1 &  0 & -1 & -1
    \end{bmatrix}\,,
\end{equation*}
the generating function is then :
\begin{equation*}
    \begin{split}
        G'_6(X,Y_0,Y_1)
        &= \prod_{k=0}^5 \sum_{n_k=0}^1 X^{n_k} \: Y_0^{I^6_0(n_k)} \: Y_1^{I^6_1(n_k)} \\
        &= \left[1 + X Y_0\right] \left[1 + X Y_1\right] \left[1 + X Y_0^{-1} Y_1\right] \left[1 + X Y_0^{-1}\right] \left[1 + X Y_1^{-1}\right] \left[1 + X Y_0 Y_1^{-1}\right] \\
        &= 1 + X [ Y_0^{-1} ( 1 + Y_1 ) 
            + ( Y_1^{-1} + Y_1 ) 
            + Y_0 ( Y_1^{-1} + 1 ] \\
        &\hspace{6.5mm}+ X^2 [ 3 + Y_0^{-2} Y_1  
            + Y_0^{-1} ( Y_1^{-1} + 1 + Y_1 + Y_1^2 ) 
            + ( Y_1^{-1} + Y_1)+ Y_0 ( Y_1^{-1} + 1 + Y_1 )
            + Y_0^2  Y_1^{-1} ] \\
        &\hspace{6.5mm}+ \mathcal{O}(X^3)\,,
    \end{split}
\end{equation*}
where the single $X^0$ term correspond to the only $N=0$ configuration : $[0,0,0,0,0,0]$, and the $C_{1}(6,1)=6$ number of $X^1$ terms correspond to the $N=1$ configurations,
i.e. $[1,0,0,0,0,0]$ and its cyclic permutations.
The $C_{1}(6,2)=15$ number of $X^2$ terms correspond to the $N=2$ configurations we are interested in here. From here we can either directly read-off the coefficients from the polynomial expansion, or take the formal approach of applying consecutive partial derivatives below. 

It is of note that for degeneracy classes that are specified by negative quantum numbers $I^q_k<0$ the partial derivative approach requires a change of notation.

Taking first the partial derivative with respect to $X$ to select the $N = 2$ configurations :
\begin{equation*}
\begin{split}
    \frac{1}{2!} \left. \frac{\partial^2}{\partial X^2} \right\rvert_{X=0} G'_6(X,Y_0,Y_1) &= 3 + Y_0^{-2} Y_1  
            + Y_0^{-1} ( Y_1^{-1} + 1 + Y_1 + Y_1^2 ) 
            + ( Y_1^{-1} + Y_1)\\&\hspace{6.5mm}+ Y_0 ( Y_1^{-1} + 1 + Y_1  + Y_1^{-2})
            + Y_0^2  Y_1^{-1}\,.
\end{split}
\label{firstdiff}
\end{equation*}

We then apply the partial derivatives with respect to $Y_0$ and $Y_1$ to select the $U^L_\ell(n) = 0$ degeneracy class which has invariant values $I^6_0 = I^6_1 = 0$ :
\begin{equation*}
    \frac{1}{2!} \left. \frac{\partial^2}{\partial X^2} \right\rvert_{X=0}
    \frac{1}{0!} \left. \frac{\partial^{I^6_0=0}}{\partial Y_0^{I^6_0=0}} \right\rvert_{Y_0=0}
    \frac{1}{0!} \left. \frac{\partial^{I^6_1=0}}{\partial Y_1^{I^6_1=0}} \right\rvert_{Y_1=0} G'_6(X,Y_0,Y_1) = 3\,,
\end{equation*}
which corresponds to the degeneracy class at the origin containing $3$ configurations. This is exactly the result we expected since direct numerical
computations confirm that those $3$ configurations are indeed $[1,0,0,1,0,0]$ and its cyclic permutations.

Alternatively, we could instead select the $U^L_\ell(n) = 1$ degeneracy class which has invariant values $I^6_0 = 1, I^6_1 = 0$ and thus we compute :
\begin{equation*}
    \frac{1}{2!} \left. \frac{\partial^2}{\partial X^2} \right\rvert_{X=0}
    \frac{1}{1!} \left. \frac{\partial^{I^6_0=1}}{\partial Y_0^{I^6_0=1}} \right\rvert_{Y_0=0}
    \frac{1}{0!} \left. \frac{\partial^{I^6_0=1}}{\partial Y_1^{I^6_1=0}} \right\rvert_{Y_1=0} G'_6(X,Y_0,Y_1) = 1\,,
\end{equation*}
which corresponds to the degeneracy class at $(1,0)$ in the complex plane containing one configuration. Once again, direct numerical computations
confirm this is the correct result, the single configuration being $[0,1,0,0,0,1]$.

\subsection{Example : Bosons ($R = N$), $L=6, N=3, \ell=2$}

Let us now look at $N = 3$ using a different sector : when $\ell = 2$ we get the $q = L/\ell = 3$ sector. We are now working with $3$-roots of unity so
we use the previous section's induction method to compute the $T_3$ matrix :
\begin{equation*}
    T_3^\top = \begin{bmatrix}
        1 & 0 & -1 \\
        0 & 1 & -1
    \end{bmatrix}\,.
\end{equation*}

The generating function for the $q=3$-sector is then :
\begin{equation*}
\begin{split}
    G'_3(X,Y_0,Y_1)
        &= \prod_{k=0}^5 \sum_{n_k=0}^3 X^{n_k} \: Y_0^{I_0^3(n_k)} \: Y_1^{I^3_1(n_k)} \\
        &= \left( \left[1 + X Y_0 + X^2 Y_0^2 + X^3 Y_0^3\right] \left[1 + X Y_1 + X^2 Y_1^2 + X^3 Y_1^3\right] \right. \\
        &\hspace{5mm} \left. \left[1 + X Y_0^{-1} Y_1^{-1} + X^2 Y_0^{-2} Y_1^{-2} + X^3 Y_0^{-3} Y_1^{-3}\right] \right)^2 \\
    &=1+2X(Y_0+Y_1+Y_0^{-1}Y_1^{-1})+X^2(3Y_0^2+3Y_1^2+4Y_0Y_1+3Y_0^{-2}Y_1^{-2}+4Y_1^{-1}+4Y_0^{-1})\\
    &\hspace{6.5mm}+X^3(4Y_0^3+4Y_1^3+6Y_0^2Y_1+6Y_0Y_1^2+4Y_0^{-3}Y_1^{-3}+6Y_0Y_1^{-1}+6Y_0^{-1}Y_1+8\\
    &\hspace{17mm}+6Y_0^{-1}Y_1^{-2}+6Y_0^{-2}Y_1^{-1})\,.
\end{split}
\end{equation*}

Taking first the partial derivative with respect to $X^3$ to restrict the compositions to $N=3$, we find:
\begin{equation}
\begin{split}
    \frac{1}{3!}\left.\frac{\partial^3}{\partial X^3}\right\rvert_{X=0}G'_{q=3}(X,Y_0,Y_1)=(&4Y_0^3+4Y_1^3+6Y_0^2Y_1+6Y_0Y_1^2+4Y_0^{-3}Y_1^{-3}+6Y_0Y_1^{-1}\\&+6Y_0^{-1}Y_1+8+6Y_0^{-1}Y_1^{-2}+6Y_0^{-2}Y_1^{-1})\,,
\end{split}
\label{eq:L6N3l2ModGen}
\end{equation}

We can compare \cref{eq:L6N3l2ModGen} to the $U^L_l$ distribution (\cref{fig:UL6N3l2}) and see that the degeneracies of all the terms matches
those of the distribution by either performing the partial derivatives with respect to $Y_0$ and $Y_1$, or by examining the terms and the
associated coefficients in \cref{eq:L6N3l2ModGen} directly.

\begin{figure}[!ht]
    \centering
    \includegraphics[width=0.8\linewidth]{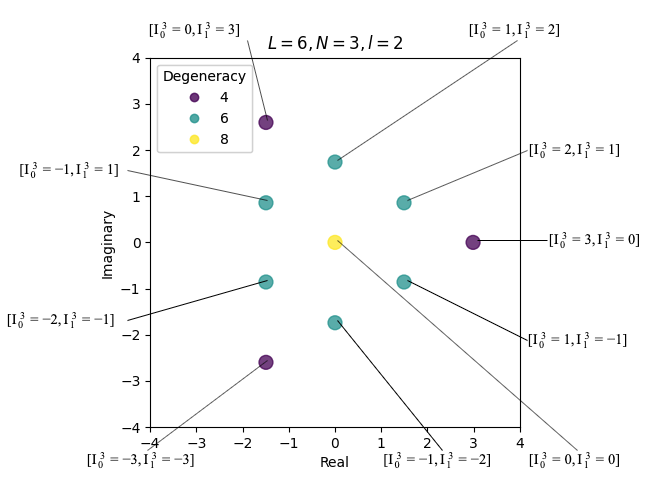}
    \caption{\textbf{Illustration of the $L=6,N=3,l=2$ $U^L_l$ distribution, the degeneracy classes are color coded to denote their multiplicity,
    and labeled with the quantum number values.}
    These degeneracy values and invariant values can be compared to the terms in \cref{eq:L6N3l2ModGen}, where the $I_0^3, I_1^3$ values are
    encoded as the exponents of $Y_0$ and $Y_1$ respectively, and the degeneracy is the integer value of the coefficients associated with the terms.}
    \label{fig:UL6N3l2}
\end{figure}
\end{widetext}

\section{Many-body Energy Resummation}
\label{appendix:e}

To perform the resummation of the many-body energies, we start from \cref{resum} and decompose the sum into $q$-sectors of the lifting map :
\begin{align*}
    E &= \sum_{\ell=0}^{L-1} U^L_\ell \: \tilde\epsilon^L_\ell \,, \\
        &= N \tilde\epsilon^L_0 + \sum_{\underset{\ell|L,\ell<L}{q=L/\ell}} \: \sum_{\underset{\gcd(L,p')=\gcd(L,\ell)}{p'<L}} U^L_{p'} \: \tilde\epsilon^L_{p'} \,,
\end{align*}
where we introduced the outer sum over all $q$-sectors and the inner sum over the members of each $q$-sector. We then notice that $\ell$ must
divide $p'$ so we perform a change of variable $p' = p\ell$ in the inner sum to obtain the reduced distributions :
\begin{equation*}
    E = N \tilde\epsilon^L_0 + \sum_{\underset{\ell|L,\ell<L}{q=L/\ell}} \: \sum_{\underset{\gcd(L,p)=\gcd(L,\ell)}{p\ell<L}} U^q_{p} \: \tilde\epsilon^q_{p} \,.
\end{equation*}
At this point we want to switch notation and express the value of a single many-body energy, labeled by occupation numbers :
\begin{equation*}
    E_n = N \tilde\epsilon^L_0 + \sum_{\underset{\ell|L,\ell<L}{q=L/\ell}} \: \sum_{\underset{\gcd(L,p)=\gcd(L,\ell)}{p\ell<L}} U^q_{p}(n) \: \tilde\epsilon^q_{p} \,.
\end{equation*}
The next step is to inject the expansion of $U^q_{p}(n)$ over the cyclotomic basis in $\mathbb{Q}(\omega_q)$ for each $q$-sector, we get :
\begin{equation*}
    E_{\{I^q_{k'}\}} = N \tilde\epsilon^L_0 + \sum_{\underset{\ell|L,\ell<L}{q=L/\ell}} \: \sum_{\underset{\gcd(L,p)=\gcd(L,\ell)}{p\ell<L}} \left(
        \sum_{k'=0}^{\varphi(q)-1} I^q_{k'} \: \omega_q^{pk'}
    \right) \: \tilde\epsilon^q_{p} \,,
\end{equation*}
where the many-body energy is now indexed by the quantum numbers $\{I^q_{k'}\}$ instead of the occupation numbers. We then rewrite
$\omega_q^{pk'}$ using  Froebenius automorphisms explicitly :
\begin{equation*}
    E_{\{I^q_{k'}\}} = N \tilde\epsilon^L_0 + \sum_{\underset{\ell|L,\ell<L}{q=L/\ell}} \: \sum_{\underset{\gcd(L,p)=\gcd(L,\ell)}{p\ell<L}} \left(
        \sum_{k'=0}^{\varphi(q)-1} I^q_{k'} \: \sigma^q_p(\omega_q^{k'})
    \right) \: \tilde\epsilon^q_{p} \,.
\end{equation*}
We can now re-interpret the second sum which is simply enumerating the Galois group elements, simplifying our notation, we obtain :
\begin{equation*}
    E_{\{I^q_{k'}\}} = N \tilde\epsilon^L_0 + \sum_{\underset{\ell|L,\ell<L}{q=L/\ell}} \: \sum_{\sigma^q_p \in G_q} \left(
        \sum_{k'=0}^{\varphi(q)-1} I^q_{k'} \: \sigma^q_p(\omega_q^{k'})
    \right) \: \tilde\epsilon^q_{p} \,,
\end{equation*}
which is exactly the result of \cref{eq:resum}.

\clearpage
\section{Lower Bound on the Heuristic Function $h_q(\epsilon)$}
\label{appendix:lowerbound}

We defined $T_q$ matrices such that they transform the cyclotomic basis $\mathcal{B}_q$ of a $q$-sector into the full set of
$q$-roots of unity. Applying a Froebenius automorphisms $\sigma^q_p \in G_q$ to the definition of $T_q$ yields :
\begin{align*}
    V^q_1 &= T_q \; \mathcal{B}_q \,, \\
    \implies V^q_p = \sigma^q_p(V^q_1) &= T_q \; (\sigma^q_p(\mathcal{B}_q)) = T_q \; \Omega^q_p \,,
\end{align*}
where we defined :
\begin{equation*}
    \Omega^q_p = \sigma^q_p(\mathcal{B}_q) = \{ \omega_q^0, \omega_q^p, \omega_q^{2p}, \dots, \omega_q^{p(\varphi(q)-1)} \} \,.
\end{equation*}
This allows us to rewrite the effective single-body energies associated to a $q$-sector as follows :
\begin{align*}
    \tilde\epsilon^q_p &= (T_q \Omega^q_p)^\dagger \cdot \epsilon^q \\
        &= ( (\Omega^q_p)^\dagger T^\top_q ) \cdot \epsilon^q \\
        &= (\Omega^q_p)^* \cdot (T^\top_q \: \epsilon^q) \,.
\end{align*}
where $\epsilon^q$ is the folded vector of single-body energies in the $q$-sector defined for $q=L/\ell$ with components :
\begin{equation*}
    \forall k \in \{0,\dots,q-1\}, \: \epsilon^q_k = \sum_{p=0}^{\ell-1} \epsilon_{pq+k} \,.
\end{equation*}

We saw before that $T_q$ matrices have matrix elements which are coefficients of the cyclotomic polynomials : up to $q=105$,
those are all $\pm1$ or $0$ ($105 = 3*5*7$ is the first number to be a product of three distinct odd primes). For our purpose,
we can consider these coefficients to be of order $1$ (i.e. less than $10$ distinct odd primes in the decomposition of $q$,
which is roughly $q < 10^{11}$). It follows that the $T_q^\top \: \epsilon$ vector simply contains sums and differences of
$\epsilon_k$ values.

On the other hand, we defined $\Omega^q_p$ with entries that are $q$-roots of unity. And thus, from the above equation,
we deduce that $\tilde\epsilon^q_p$ values can be written as a linear combination of $q$-roots of unity where the coefficients
are sums and differences of $\epsilon_k$ values. As such, we can provide a lower bound on those coefficients by using the
single-body mean level spacing $\Delta_{SB}$ of the single-body spectrum.

It follows that we can give an estimate of the lower bound on the heuristic function $h_q(\epsilon)$ as :
\begin{align*}
    \min_{\tau \in S_L}[h_q(\tau[\epsilon])]
        &\sim \frac{\varphi(q) \times \min_{\gcd(q,p)=1}[|\tilde\epsilon^q_p|^2]}{L \sum_{k=0}^{L-1}|\epsilon_k|^2} \,, \\
        &\sim \frac{\varphi(q) \times \min_{\gcd(q,p)=1}[|(\Omega^q_p)^* \cdot (T^\top_q \: \epsilon^q)|^2]}{L \sum_{k=0}^{L-1}|\epsilon_k|^2} \,, \\
        &\sim \frac{\varphi(q) \times \min_{\gcd(q,p)=1}[|(\Omega^q_p)^* \cdot \Delta_{SB} \vec{1} |^2]}{L \sum_{k=0}^{L-1}|\epsilon_k|^2} \,, \\
        &\sim \frac{\varphi(q) \times \Delta^2_{SB} \min_{\gcd(q,p)=1}[|(\Omega^q_p)^* \cdot \vec{1} |^2]}{L \sum_{k=0}^{L-1}|\epsilon_k|^2} \,, \\
        &\sim \frac{\varphi(q) \times \Delta^2_{SB} \min_{\gcd(q,p)=1}[|\sum_{k=0}^{\varphi(q)-1} \omega_q^{pk}|^2]}{L \sum_{k=0}^{L-1}|\epsilon_k|^2} \,, \\
        &\sim \frac{\varphi(q)^2 \times \Delta^2_{SB}}{L \sum_{k=0}^{L-1}|\epsilon_k|^2} \,,
\end{align*}
which is the result given in \cref{eq:heuristicfunc}.

\end{document}